\documentclass[showpacs,preprintnumbers,amsmath,amssymb]{revtex4}
      \newcommand{\beq}{\begin{equation}}
      \newcommand{\eeq}{\end{equation}}
      \newcommand{\beqa}{\begin{eqnarray}}
      \newcommand{\eeqa}{\end{eqnarray}}
      
      \newcommand{\nn}{\nonumber}

      \newcommand{\del}{\partial}

    \renewcommand{\(}{\left(}
    \renewcommand{\)}{\right)}

      \newcommand{\al}{\alpha}
      \newcommand{\be}{\beta}
      
      \newcommand{\de}{\delta}
      \newcommand{\ep}{\epsilon}
      
      \newcommand{\la}{\lambda}

     \newcommand{\bbe}{\mbox{\boldmath $\beta$}}
     
      \newcommand{\ba}{\mbox{\boldmath $a$}}
     \newcommand{\bg}{\mbox{\boldmath $g$}}
     \newcommand{\bc}{\mbox{\boldmath $c$}}
     \newcommand{\bee}{\mbox{\boldmath $e$}}

     \newcommand{\bV}{\mbox{\boldmath $V$}}

\usepackage{graphicx}
\usepackage{dcolumn}
\usepackage{bm}
\begin{document}
\preprint{}
\title
{
 Deformation of a renormalization-group equation \\
 applied to infinite-order phase transitions  }
 
 \author{Hisamitsu Mukaida}
 \email{mukaida@saitama-med.ac.jp}
 \affiliation{Department of Physics, Saitama Medical College, 
 981 Kawakado, Iruma-gun, Saitama, 350-0496, Japan
 }

\date{23 August 2003\\
  ,,Revised: 04 December 2003}

\begin{abstract}
By adding a linear term to a renormalization-group equation in a 
system exhibiting infinite-order phase transitions,  asymptotic 
behavior of running coupling constants is derived in 
an algebraic manner.   A benefit of this method is 
presented explicitly using several examples. 
\end{abstract}


\maketitle

\section{Introduction}
Renormalization-group (RG) technique is  one of the most powerful methods for 
investigating critical phenomena in statistical physics\cite{wk}.  
In general,  RG transformation (RGT)
consists of a coarse graining and a rescaling. It reduces   
many-body effects in a statistical model to an ordinary differential equation
of  coupling constants. 
The differential equation is called  the RG equation (RGE), and has 
generally the following form:
\beq
  \frac{d \bg}{dt} = \bV\(\bg\), 
\label{orgRGE}
\eeq
where  $\bg=(g_1,\dots, g_n)$,  a collection of coupling constants 
depending on $t$, and 
$t=\log L$ with $L$ giving the length scale of the coarse graining
 in the RG.  
One obtains a beta function  $\bV\(\bg\) = (V_1(\bg),\dots, V_n(\bg))$ 
by applying the RGT explicitly to a statistical model. 
We can derive universal exponents that characterize 
critical phenomena from asymptotic behavior 
of  solutions of Eq. (\ref{orgRGE})  for large $t$.  

\begin{figure}
\begin{center}
\setlength{\unitlength}{1mm}
\begin{picture}(80, 33)(0,0)
        \put(0,0){ 
		\includegraphics[height=33mm]{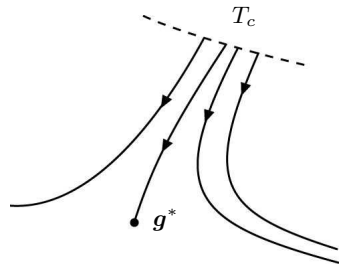}
		}
 				\put(30.5,32){$T_{c}$}
				\put(20,5){$\bg^{*}$}
\end{picture}
\end{center}
\caption{Typical RG trajectories near a phase transition. As $T$ changes, 
an initial value moves on the dashed line.  The trajectory with  $T=T_{c}$ is
absorbed into the fixed point. }
\label{fig_critical}
\end{figure}

Since the asymptotic behavior 
 is determined by vicinity of a fixed point $\bg^*$, linearization of 
 $\bV\(\bg\)$ about $\bg^{*}$ is  effective enough to obtain the 
 exponents. 
 For example, in a second-order phase
 transition,   the correlation length $\xi$ typically behaves 
 as  
\beq
  \xi = {\rm const.}/|T-T_c|^{\nu}, 
\label{2nd}
\eeq
where  $\nu$ is the correlation-length exponent and 
$T$ is a parameter specifying a state in a statistical model 
(e.g., the temperature). 
In the language of RG, $T$ parametrizes  initial values of RGE.  The 
trajectory starting from the initial value at $T=T_{c}$ is absorbed
into the fixed point.  Other trajectories  approach $\bg^{*}$ once
 but leave the fixed point subsequently, as shown in Fig. \ref{fig_critical}. 
This implies that it takes longer  for $\bg$ to leave the fixed point as 
$T$ approaches $T_{c}$.  If the scaling matrix 
$M\(\bg^{*}\)$, 
where
\beq
  M_{i j}\(\bg^*\) \equiv \frac{\del V_i}{\del g_j}\(\bg^*\)
\eeq
has a unique positive eigenvalue $\al$,  then this period  behaves as 
${\rm const. } \times (T-T_{c})^{-\al}$.  The universal exponent 
$\nu$ is obtained from $\al$ by $\nu = 1/\al$. Thus   
we do not need to find an explicit solution of  generally 
nonlinear RGE (\ref{orgRGE}). 

On the other hand, in the case of infinite-order  phase transitions, 
 $\xi$ behaves as 
\beq
  \xi = {\rm const.}  \times \exp\(A/|T-T_c|^\sigma\),
\label{infinite} 
\eeq
where $\sigma$ is a universal exponent and 
$A$ is a non-universal constant. 
Such behavior is observed when all the coupling constants are 
marginal, i.e., the canonical dimensions of the coupling constants 
are zero at $\bg^{*}$.   
Since  the linear term in $\bV\(\bg\)$ is proportional to 
the canonical dimensions of $\bg$, 
 $M_{ij} (\bg^*) = 0$ for all $i$ and $j$.  
 It indicates that we cannot extract the asymptotic behavior 
from  the scaling matrix $M(\bg^{*})$ in an infinite-order  phase transition, 
in contrast to a second-order one.  Therefore,  explicit 
solutions were traditionally required in the case of 
 an infinite-order phase transition such as the BKT phase transition\cite{bkt}.

This difficulty has been recently overcome in  Ref.\cite{im}, 
where an  RG  for RGE  (\ref{orgRGE})  is used for 
deriving asymptotic behavior of solutions.  
A general idea of  RG,   applied as a tool for asymptotic analysis 
of  non-linear differential equations,  is developed in 
Refs.\cite{cgo,bk}.  

In this report, 
we present another method. Namely, 
we derive $\sigma$  from the following deformed RGE:
\beq
  \frac{d \bg}{dt} = \epsilon \(\bg - \bg^{*}\)+
   \bV\(\bg\) \equiv \bar{\bV}\(\bg\),  
  \label{deRGE}
\eeq
where $\epsilon$ is a real number but not necessarily small. 
As we will see in the next section, the RG equation (\ref{newRGE})  for the 
RGE (\ref{orgRGE}) has a complicated form compared with 
the deformed RGE.  Hence, using the deformed RGE makes 
derivation of the critical exponent simple. 
Another  benefit of this approach is as follows:  
suppose that an infinite-order phase transition occurs 
when the spatial dimensions of the original statistical model are $d_c$.  
Then,  the deformed RGE  can be derived when they are  $d_{c}-\ep$,  under the 
  condition that all the coupling constants have a common canonical dimension. 
This condition is satisfied by various  field-theoretical models, e.g., 
an effective theory of antiferromagnets\cite{hjssw}, 
a model containing several gauge fields\cite{wls},  
a  model describing true self-avoiding random walks\cite{p}, 
and a model of nematic elastomers\cite{xr}. 
In Ref.\cite{xr},  infrared asymptotic behavior  
in $d_c$ dimensions and that in $d_{c} - \epsilon$ dimensions 
are  analyzed separately because of  
the problem of the vanishing scaling matrix explained above.  
Our method enables us to obtain universal quantities in  
both of the cases simultaneously.   We will show this advantage in the last 
example of Sec.\ref{example}.  

\section{RGE for RGE}
Here we  summarize the results of Ref.\cite{im} that will 
be used later. We consider an RGE (\ref{orgRGE}) 
for infinite-order phase transitions that  are 
controlled by a fixed point $\bg^{*}$. 
In what follows,  we put $\bg^{*} = \bf 0$ for convenience.
Suppose that we have obtained  
 $\bV(\bg)$ by the lowest-order perturbation. 
Since  linear terms vanish in infinite-order phase transitions,  
components of $\bV(\bg)$ are quadratic  in $\bg$.  
Hence the scaling property
\beq
  \bV\(k \bg\) = k^2 \bV(\bg) 
\label{sV}
\eeq
holds in this case. 
The algebraic method to compute $\sigma$ shown in Ref.\cite{im} 
employs another RG to Eq. (\ref{orgRGE}), which   
is defined as follows:  let $\bg(t, \ba_0)$ be the solution 
of Eq. (\ref{orgRGE}) 
with the initial condition $\bg(0, \ba_0)=\ba_0$.  
Choose a real number $\tau$ and  we evolve $\bg$ in time 
by $s(\tau)$,  such that $e^\tau \bg(s(\tau), \ba_0) \in S$, where 
$S$ is a sphere with the radius $a_0 \equiv |\ba_0| $ and with the center at the 
origin.  Thus we have the map $R_\tau: \ba_0 \mapsto e^\tau \bg(s(\tau), \ba_0) 
\equiv \ba\(\tau\)$.  Thanks to Eq. (\ref{sV}), $R_\tau$ satisfies 
the semigroup property $R_{\tau+\tau'} = 
R_\tau  R_{\tau'}$, so that $R_\tau$ is called RG for RGE (\ref{orgRGE}).  
The infinitesimal transformation leads to the following new RGE:
\beq
  \frac{d \ba}{d\tau} = \bbe\(\ba\) \equiv - 
  \frac{P\(\ba\)\bV\(\ba\)}{\ba\cdot \bV\(\ba\)}a_0^2, 
\label{newRGE}
\eeq
where $P\(\ba\)$ is the projection operator defined by $P_{i j}\(\ba\) = \de_{i j} - a_i a_j/a_{0}^{2}$.  
Since $\bbe\(\ba\)$ is perpendicular to $\ba$,  solutions of the new RGE are  
restricted on $S$.  Introducing the polar coordinates $\{\theta_\al \}_{1\leq\al\leq n-1}$ on $S$
and the corresponding orthonormal basis,  
\beqa
  && \tilde{\bee}_\al \equiv f_\al\(\ba\)^{-1} \, \frac{\del \ba}{\del \theta_\al}, \ 
  f_\al\(\ba\) \equiv \left| \frac{\del \ba}{\del \theta_\al} \right|,  
\eeqa
we can expand $\bbe(\ba)$ as
\beq
  \bbe\(\ba\) = \sum_{\al =1}^{n-1} \tilde\be_\al\(\ba\) \tilde{\bee}_\al. 
\eeq
The new RGE is represented as 
\beqa
  \frac{d \theta_\al}{d \tau} \(\ba\)  = f_\al^{-1}\(\ba\) \tilde \be_\al \(\ba \) 
\eeqa
by  the polar coordinates. 

\begin{figure}
\begin{center}
\setlength{\unitlength}{1mm}
\begin{picture}(80, 32)(0,0)
        \put(0,0){ 
		\includegraphics[height=32mm]{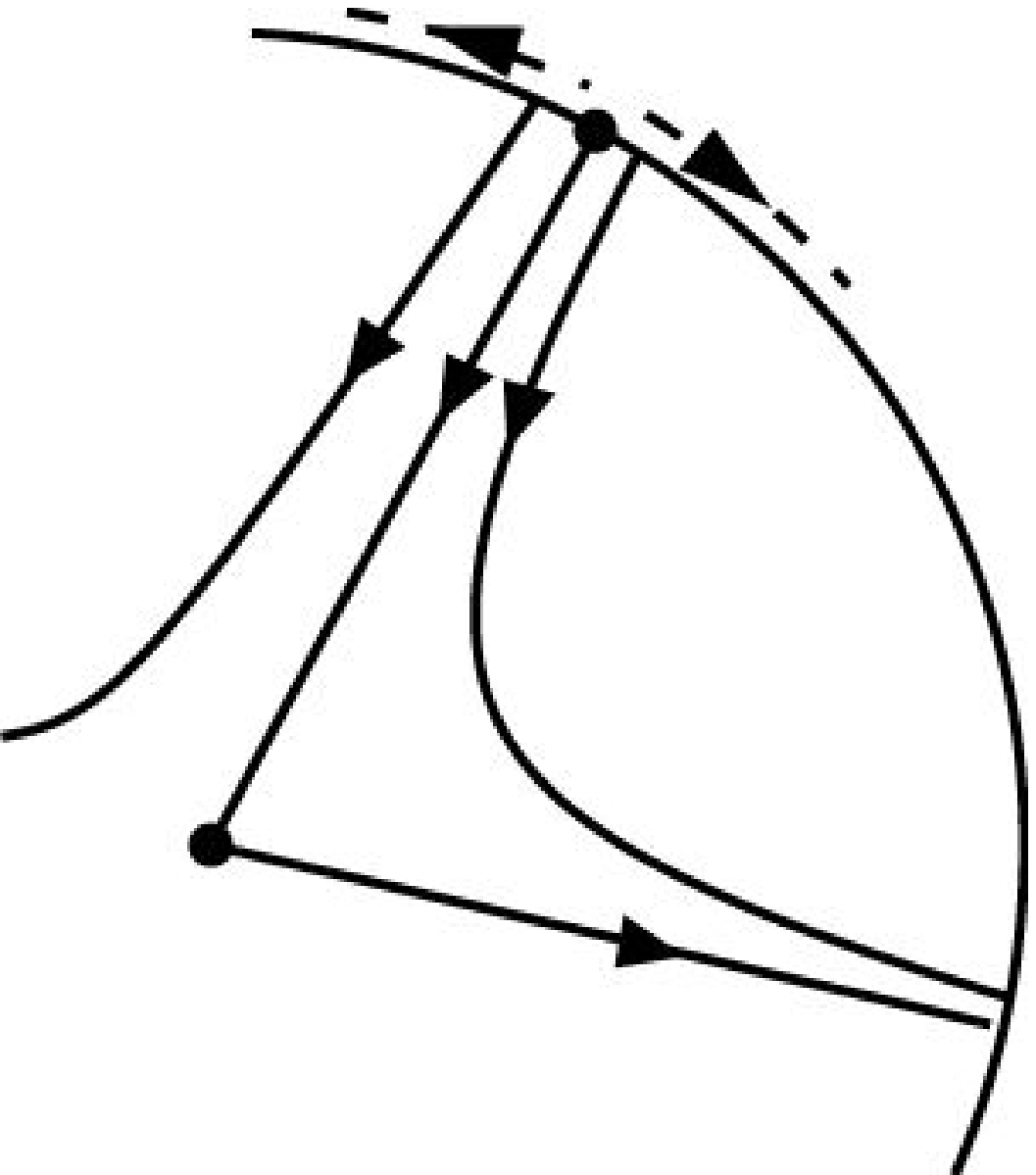}
		}
        \put(53,0){
                  \includegraphics[height =32mm]{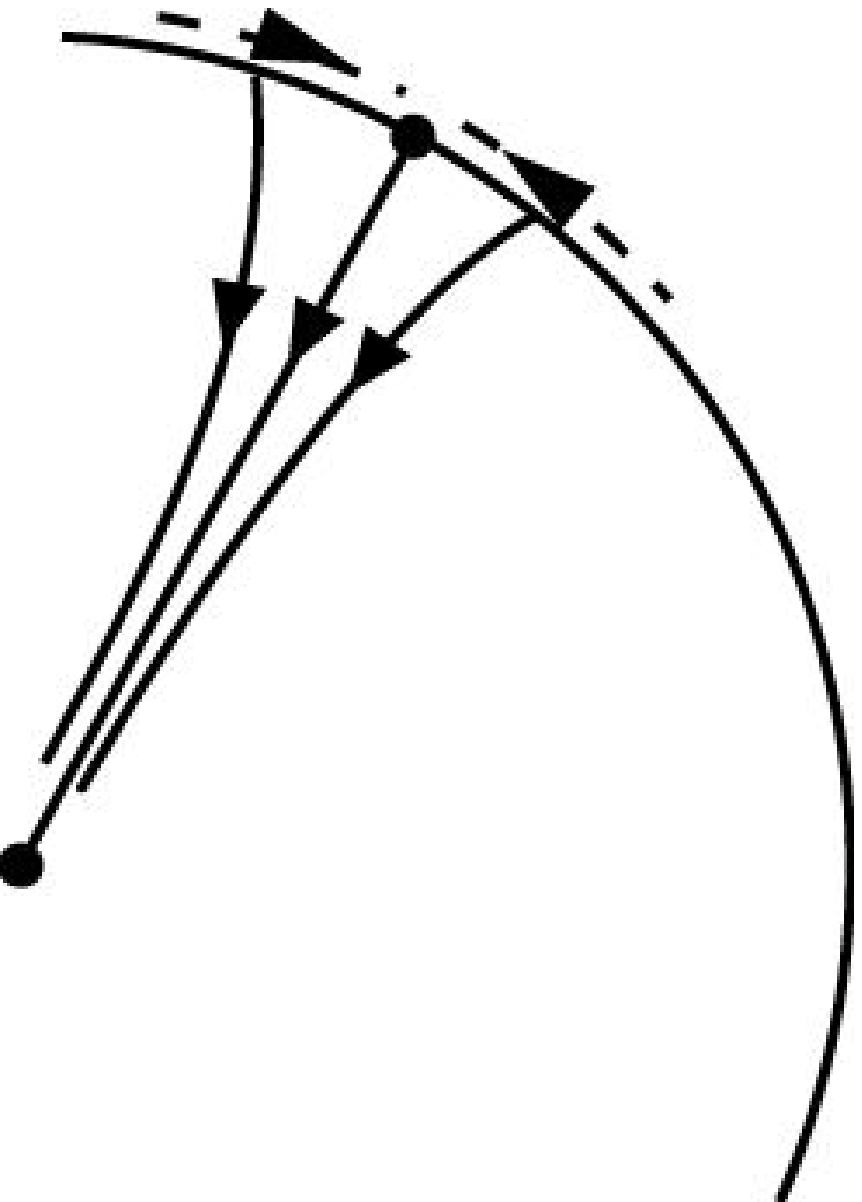}
                  }
				\put(19,33){$\ba^{*}$}
				\put(5,6){O}
				\put(8,-3){(a)}
				\put(32,0){$S$}
                                      \put(66,33){$\ba^{*}$}
                                     \put(61,-3){(b)}
				\put(53,6){O}
				\put(78,0){$S$}
\end{picture}
\end{center}
\caption{Schematic trajectories of RGE. The solid lines are for the 
original RGE (\ref{orgRGE}) while the dashed lines are for 
the new RGE (\ref{newRGE}) defined on $S$.  
Here (a) is the case where 
a unique positive eigenvalue exists in $\mu(\ba^{*})$. 
(b) is the case where all the eigenvalues of $\mu(\ba^{*})$ are negative. 
}
\label{fig_flow}
\end{figure}

It is easily found that $\ba^* \in S$ is a fixed  point of  the new RGE (\ref{newRGE})
if $\bg(t, \ba^*)$ is a straight flow line. 
In  particular, a fixed point on 
an incoming straight flow line satisfying $\ba^* \cdot \bV(\ba^*) < 0$
plays an important role because trajectories near this fixed point 
correspond to trajectories of Eq. (\ref{orgRGE}) approaching $\bg^{*}$. 
 In contrast to the original RGE,   we can generally linearize the new RGE about 
$\ba^*$.  
In Ref.\cite{im}, it is shown that  the scaling matrix of the new RGE 
\beq
  \mu_{\al \be}\(\ba^*\)  \equiv f_\al^{-1}\(\ba^*\) 
  \frac{\del \tilde{\be}_\al}{\del \theta_\be} \(\ba^* \)
\label{mu}
\eeq
plays a similar role to $M(\bg^{*})$ in the original RGE 
describing a second-order phase transition. 
 Namely, if the matrix $\mu(\ba^{*})$ has 
a unique positive eigenvalue $\la$,  in which typical trajectories of the 
original RGE  are  in Fig. \ref{fig_flow} (a),  
 we can observe divergence of the correlation length by one-parameter tuning and
\beq
  \sigma = \frac{1}{\lambda}
 \label{max}
\eeq
in Eq. (\ref{infinite}).  
On the other hand, if all the eigenvalues of $\mu\(\ba^*\)$ are negative, 
where typical trajectories are in Fig. \ref{fig_flow} (b), 
 $\bg(t, \ba_0)$
behaves as 
\beq
  \bg\(t, \ba_0\) \sim \frac{1}{C\(\ba^*\) t} \bee^*. 
\label{massless}
\eeq 
In this formula,   $\bee^* \equiv \ba^*/a_0$ and 
$C\(\bg\)$ is defined by the relation 
\beq
  C\(\bg\) |\bg|^3 =  - \bg \cdot \bV\(\bg\). 
\eeq
The asymptotic behavior in Eq. (\ref{massless}) is important for 
investigating finite size scaling in a statistical system, for example.

\section{Deformed RGE}
\begin{figure}
\begin{center}
\setlength{\unitlength}{1mm}
\begin{picture}(80, 30)(0,0)
        \put(0,0){ 
		\includegraphics[height=30mm]{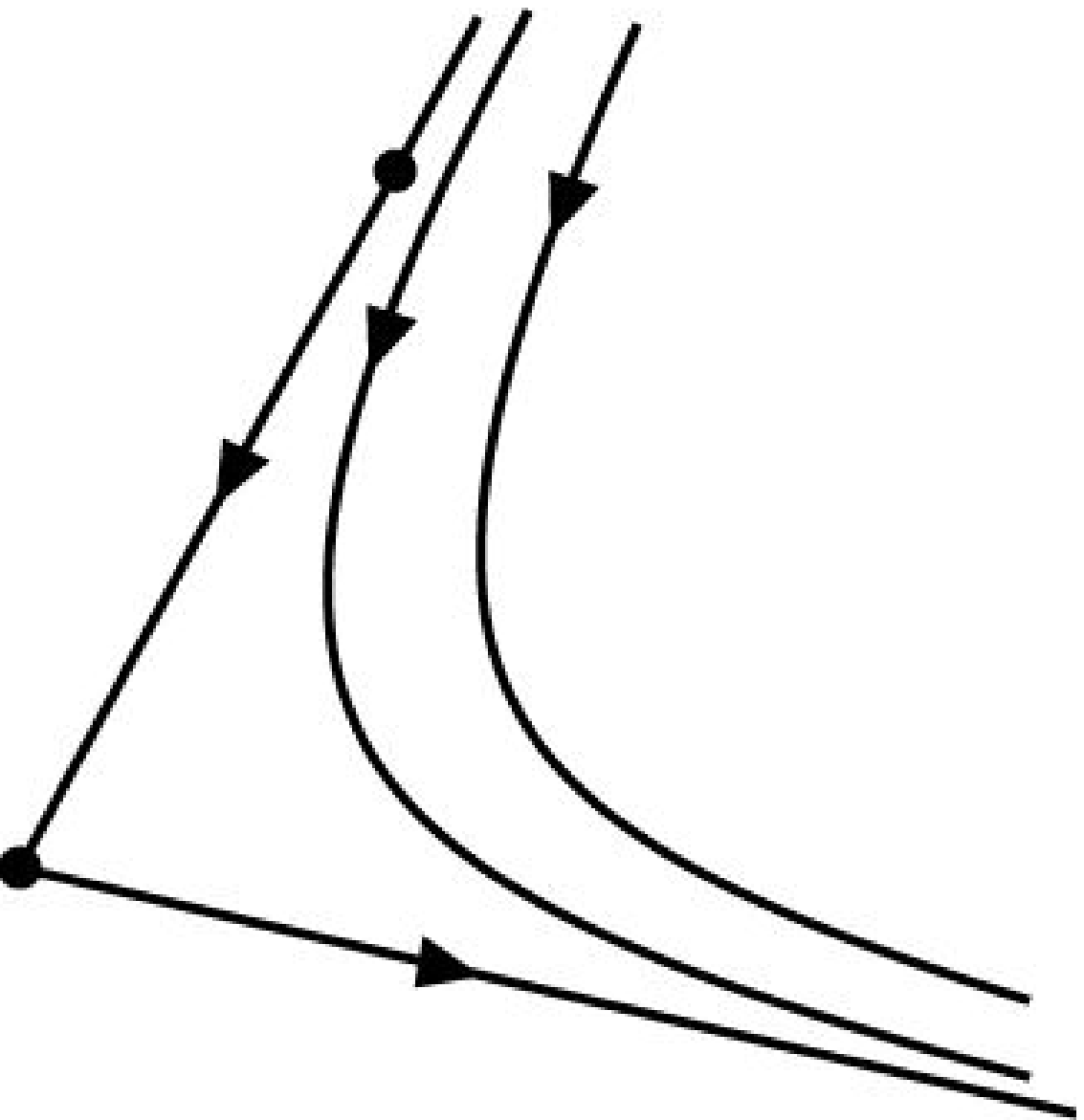}
		}
        \put(40,0){
                  \includegraphics[height =30mm]{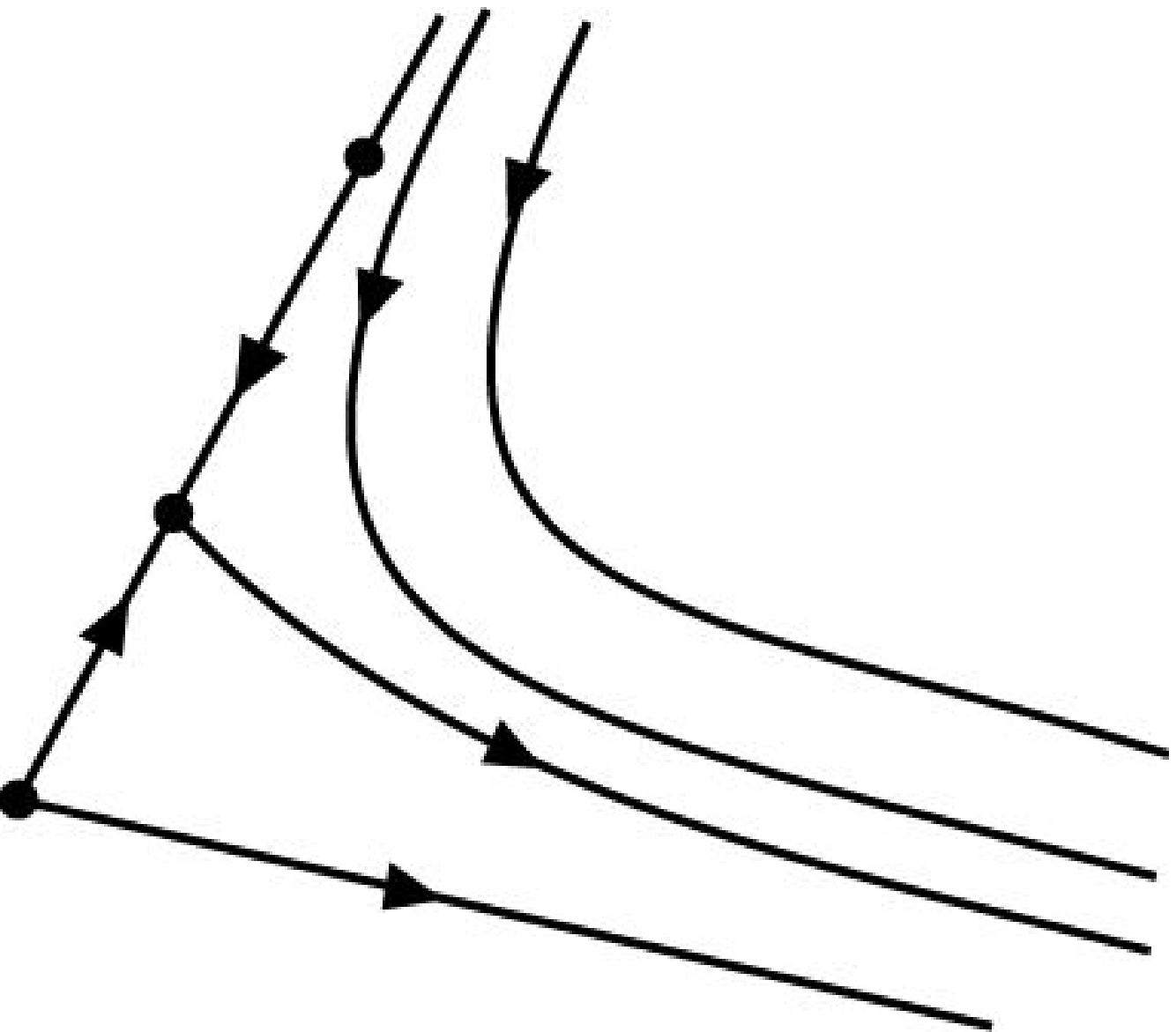}
                  }
				\put(8,25){$\ba^{*}$}
				\put(0,2){O}
				\put(10,-3){(a)}
                                      \put(48,25){$\ba^{*}$}
                                      \put(43,15){$\bc^{*}$}
                                     \put(50,-3){(b)}
				\put(40,2){O}
\end{picture}
\end{center}
\caption{(a) Schematic trajectories for the original RGE.   (b) Those for the deformed RGE. }
\label{fig_rge}
\end{figure}

Next,  we  consider the deformed RGE (\ref{deRGE}) putting 
$\bg^{*} = \bf 0$.    We can take $\ep > 0$ without loss of generality. 
A fixed point $\bc^*$ of the deformed RGE solves
\beq
   \bar{\bV}(\bc^{*}) = \epsilon \bc^* + \bV\(\bc^*\) = {\bf 0}.
\label{fxd} 
\eeq
A key feature of the deformed RGE is  that 
 $\bc^{*}$ in  Eq. (\ref{fxd}) and a fixed point $\ba^{*}$ 
 of the new RGE (\ref{newRGE}) on an 
 incoming straight flow line has  one-to-one 
 correspondence via
\beq
  \ba^* = \frac{a_0}{c^*} \bc^* 
\eeq
as depicted in Fig. \ref{fig_rge}. 
Writing $\bV\(\bg\)$ as  
\beq
  \bV\(\bg\) = \sum_{\al=1}^{n-1} \tilde{V}_\al\(\bg\) \tilde{\bee}_\al 
  + \tilde{V}_n\(\bg\) \tilde{\bee}_n, 
\eeq
where $\tilde{\bee}_{n} \equiv \bg/g$, 
we have the deformed RGE in  the polar coordinates:
\beqa
  &&\frac{d \theta_\al}{d t}\(\bg\)  = f_\al^{-1}\(\bg\) \tilde{V}_\al \( \bg\)
  \nn\\
  &&\frac{d g}{d t}\(\bg\) =\ep g + \tilde{V}_n \( \bg\).
\eeqa
Expanding the above formula about the fixed point $\bc^*$, we have the following 
scaling matrix $\bar{M}\(\bc^*\)$:
\beqa
  &&\bar{M}_{\al \be}\(\bc^*\) = 
  f_\al^{-1}\(\bc^*\)\frac{\del \tilde{V}_\al}{\del \theta_\be}\( \bc^*\) 
  \nn\\
  &&\bar{M}_{\al n}\(\bc^*\) = 
  f_\al^{-1}\(\bc^*\)\frac{\del \tilde{V}_\al}{\del g}\( \bc^*\)
  \nn\\
  &&\bar{M}_{n\al}\(\bc^*\) = \frac{\del \tilde{V}_n}{\del \theta_\be} \( \bc^*\)
  \nn\\
  &&\bar{M}_{nn}\(\bc^*\) =  \( \ep +  \frac{\del \tilde{V}_n}{\del g} \( \bc^*\)\), 
 \label{M}
\eeqa
where $\al$ and $\be$ run from 1 to $n-1$.  
Since $\tilde{V}_\al \(\bg\)$ is a component  perpendicular to $\bg$, 
one finds that $\tilde{V}_\al \(k \bc^*\) = 0$ for all $k$ with help of  
Eqs.(\ref{sV}) and (\ref{fxd}).  This means that 
$\del_g \tilde{V}_\al \(\bc^*\) = 0$.   On the other hand,  
$\del_g \tilde{V}_n\(\bc^*\) = 2 \tilde{V}_n\(\bc^*\)/g^* = -2\epsilon$  
because $\tilde{V}_n\(\bg\)$ is quadratic in $g$. Therefore, 
\beq
  \bar{M}_{\al n} = 0, \ \ \bar{M}_{nn} = -\epsilon
  \label{entry}
\eeq
in Eq. (\ref{M}).  
 Furthermore, we can rewrite $\bar{M}_{\al \be}\(\bc^{*}\)$ 
 in terms of $\mu_{\al \be}\(\ba^{*}\)$.   In fact, 
 $\mu\(\ba^{*}\)$ in  Eq. (\ref{mu}) is written as  
\beqa
  \mu_{\al \be}\(\ba^*\) &=& 
  f_\al^{-1}\(\ba^*\) \frac{1}{C\(\ba^*\)a_0}
  \frac{\del \tilde{V}_\al}{\del \theta_\be} \(\ba^* \). 
\eeqa
Employing  the following scaling properties:
\beqa
 && C\(k \bg \) = C\(\bg\) \nn\\
 && f_\al \(k \bg\) = k f_{\al} \(\bg\) \nn\\
 &&\frac{\del \tilde V_{\al}}{\del \theta_\be} \(k \bg\) = 
 k^{2} \frac{\del \tilde V_{\al}}{\del \theta_\be} \(\bg\), 
\label{scaling}
\eeqa
we get 
\beq
    \bar{M}_{\al \be} \(\bc^*\) =  \epsilon \mu_{\al \be}\(\ba^*\).  
\label{res1}
\eeq
Eqs.(\ref{entry}) and (\ref{res1}) shows that $\bar{M}\(\bc^{*}\)$ has 
a form of 
\beq
  \bar{M}\(\bc^{*}\) = 
  \(
  \begin{array}{ccc|c}
  &&&0\\
  \multispan2 \hfill $ \ep \mu\(\ba^{*}\)$  &&\vdots\\
  &&&0\\
  \hline
  *&\cdots&*&-\ep\\
  \end{array}
  \).
\eeq
in the polar coordinates.   It readily follows from this formula that 
$\bar{M}(\bc^{*}) \tilde{\bee}_{n} = -\ep \tilde{\bee}_{n}$. 
Thus we can derive all the eigenvalues of $\mu (\ba^{*})$ from $\bar{M}(\bc^{*})$ by 
removing $-\ep$, which is the eigenvalue corresponding to the 
eigenvector $\tilde{\bee}_{n}$,  
from the set of the eigenvalues of $\bar{M}(\bc^{*})$, and,  by 
multiplying by $1/\ep$,   the remaining eigenvalues. 
 Further, if  all the eigenvalues of $\bar{M}\(\bc^{*}\)$  are negative,  
$\bg(t, \ba_0)$ behaves as 
\beq
  \bg(t, \ba_0) \sim \frac{1}{C\(\ba^*\)t}\bee^* = \frac{1}{\epsilon t}\bc^*,  
\label{massless2} 
\eeq
according to Eq. (\ref{massless}) and the scaling property of $C\(\ba^{*}\)$
in Eq. (\ref{scaling}).

\section{Example}
\label{example}
Here are  several examples. 
The first example is taken from the two-dimensional classical XY model\cite{bkt}. 
Here,  the beta function $\bV\(\bg\)$ is given as 
\beq
  \bV\(\bg\) = \(
    \begin{array}{c}
    - g_2^2\\
    - g_1 g_2
    \end{array}
    \), 
\eeq
for $g_{1}, g_{2} >0$.  The deformed RGE has the fixed point $\bc^*=\(\ep, \ep\)$. 
The scaling matrix $\bar{M}\(\bc^*\)$ of the deformed RGE is easily computed 
in  terms of the cartesian coordinates as 
\beq
  \bar{M}\(\bc^*\) = \(
  \begin{array}{cc}
    \ep&-2\ep\\
    -\ep&0
  \end{array}
  \). 
\eeq
It has the eigenvalues $-\ep$ and $2\ep$.    Employing Eq. (\ref{max}), we get 
\beq
  \sigma = \frac{\ep}{2\ep} = \frac{1}{2}, 
\eeq
which is a well-known result.  As we have explained in the previous section, 
the other eigenvalue,  $-\ep$,  always appears in a deformed RGE (\ref{deRGE}), 
which corresponds to the eigenvector $\bc^{*}/c^{*}$.  

The next example is the RGE in a one-dimensional quantum spin 
chain,  studied by Itoi and Kato\cite{ik}; it is defined by 
\beq
  \bV\(\bg\) = \(
    \begin{array}{c}
    g_{1}\(Ng_{1}+2g_{2}\)\\
    -g_2\(2g_{1}+Ng_{2}\)
    \end{array}
    \). 
\eeq
The deformed RGE has the following three nontrivial fixed points:
\beq
  \bc^*_{1} = \(-\frac{\ep}{N},0\), \ 
  \bc^*_{2} = \(0,\frac{\ep}{N}\), \ 
  \bc^*_{3} = \frac{\ep}{N-2} \(-1,1\). 
\eeq
The corresponding scaling matrices are 
\beqa
  && \bar{M}_{1}=\( \begin{array}{cc}
       -\ep & -\frac{2\ep}{N} \\
       0& \frac{N+2}{N}\ep
     \end{array}
     \), \ 
     \bar{M}_{2}=\( \begin{array}{cc}
       \frac{N+2}{N}\ep & 0 \\
       -\frac{2\ep}{N}& -\ep
     \end{array}
     \),  \nn\\
  &&\bar{M}_{3}=\( \begin{array}{cc}
       \frac{N \ep}{2-N} & \frac{2 \ep}{2-N} \\
       \frac{2 \ep}{2-N}& \frac{N \ep}{2-N}
     \end{array}
     \). 
\eeqa
The eigenvalues of those matrices are,  respectively,  
\beq
  \frac{N+2}{N}\ep, \ \frac{N+2}{N}\ep, \ {\rm and } \ \frac{2+N}{2-N}\ep, 
\eeq
up to the common eigenvalue $-\ep$. 
The other eigenvalues divided by $\epsilon$ are equal to those of the scaling matrices 
derived from the new RGE (\ref{newRGE}), which is computed in Ref.\cite{im}. 
It should be noted that the deformed RGEs in the above two examples 
do not correspond  to those in $2 -\ep$ and  
$1-\ep$ dimensions respectively.   However,  the derivation presented here 
is much simpler than the method using Eq.(\ref{newRGE}).

The last example is the RGE in a field-theoretical model 
for  nematic elastomers,   proposed in Ref.\cite{xr}.  
In contrast to the previous examples, 
the deformed RGE is  obtained exactly in $3-\ep$ dimensions with 
\beq
  \bV \(\bg\) = \frac{-1}{8\(4g_{1}+g_{2}\)}
  \(
  \begin{array}{c}
    g_{1}\(40 g_{1}^{2}+68g_{1}g_{2} + 13g_{2}^{2}\)\\
    2g_{2}\(4 g_{1}^{2}+32g_{1}g_{2} + 7g_{2}^{2}\)
  \end{array}
  \). 
\label{nematic}
\eeq
Although $\bV\(\bg\)$ is not quadratic polynomial, our result is applicable
 because all we need to apply the present method is the scaling  property
 of $\bV\(\bg\)$,  Eq. (\ref{sV}).  
 The deformed RGE 
has the three fixed points
\beq
  \bc_{1}^{*} = \(\frac{4\ep}{5}, 0\), \ \ \bc_{2}^{*}=\(\frac{4\ep}{59}, \frac{32\ep}{59}\), 
  \ \ \bc_{3}^{*} = \(0,\frac{4\ep}{7}\). 
 \label{nemaFxd}
\eeq
One can check that the scaling matrices  
 have the following respective eigenvalues 
\beq
 4\ep/5,  \ \  -4\ep/59,   \ {\rm and} \  \ep/14
 \label{nemaVal}
\eeq
in addition to the common eigenvalue $-\ep$. 
Now we  turn to the case of just three dimensions. 
 If $g_{1}, g_{2} >0$, infrared behavior of a system is 
 governed by the fixed point $\bc^{*}_{2}$\cite{xr}. 
 Since the eigenvalue at $\bc^{*}_{2}$ is negative, 
 $\bg\(t, \ba_{0}\)$ behaves as 
  \beq
    \bg\(t, \ba_{0}\) \sim \frac{1}{\epsilon t} \bc^{*}_{2} = 
    \frac{1}{t} \(\frac{4}{59}, \frac{32}{59}\)
  \eeq
  for sufficiently large $t$, according to Eq. (\ref{massless2}). 
  The result is consistent with that in 
  Ref.\cite{xr}. 
\section{Summary}
We have shown how to derive asymptotic behavior 
of a solution of RGE for infinite-order phase transition, 
by adding a linear term to this RGE. This method can 
allow us to apply a result of the $\epsilon$ expansion 
to the case where $\epsilon = 0$.

\end{document}